\documentclass[12pt]{article}
\usepackage{amssymb}
\usepackage{mathrsfs}
\usepackage{bm}
\usepackage{graphicx}
\usepackage{subfigure}
\begin{document}
\begin{center}
\begin{Large}
{\bf Supersymmetric probability distributions}
\end{Large}
\vskip1truecm
S. Nicolis\footnote{E-Mail: Stam.Nicolis@lmpt.univ-tours.fr} and
A. Zerkak\footnote{E-Mail: ouahid.zerkak@gmail.com}

\vskip1truecm
{\sl CNRS--Laboratoire de Math\'ematiques et Physique Th\'eorique (UMR 7350)\\
F\'ed\'eration de Recherche ``Denis Poisson'' (FR2964)\\
D\'epartement de Physique, Universit\'e de Tours\\
Parc Grandmont, 37200 Tours}
\end{center}

\begin{abstract}
We use anticommuting variables to study probability distributions of random
variables, that are solutions of Langevin's equation. We show that the
probability density always enjoys ``worldpoint supersymmetry''. 
The partition function, however, may not. 
We find that the domain of integration can acquire 
 a boundary, that implies that the auxiliary field has a 
non-zero expectation value, signalling spontaneous supersymmetry breaking. 
This is due to the presence of ``fermionic'' zeromodes, whose
contribution cannot be cancelled by a surface term.
This we prove by an explicit calculation of the regularized partition
function, as well as by computing the moments of the auxiliary field and 
checking whether they satisfy the identities implied by Wick's theorem.
Nevertheless, supersymmetry manifests itself in the identities that are
satisfied by the moments of the scalar, whose expressions we can calculate,
for all values of the coupling constant. 
We also provide some quantitative estimates concerning the visibility
of supersymmetry breaking effects in the identities for the moments and remark
that the shape of the distribution of the auxiliary field can influence quite
strongly how easy it would be to mask them, since the expectation
value of the auxiliary field doesn't coincide with its typical value.

\end{abstract}
\newpage

\section{Introduction}
Let us consider a stochastic process, of commuting random
variables, $x(\tau)$, that satisfies the Langevin
equation 
\begin{equation}
\label{langevin}
\frac{d x(\tau)}{d\tau} = -\frac{\partial U(x(\tau))}{\partial x(\tau)} + 
\eta(\tau)
\end{equation} 
Here $\eta(\tau)$ is a Gaussian stochastic process, i.e. its correlation
functions satisfy the relations
\begin{equation}
\label{gaussian}
\begin{array}{l}
\displaystyle
\left\langle\eta(\tau)\right\rangle = 0\\
\displaystyle
\left\langle\eta(\tau_1)\eta(\tau_2)\right\rangle = \delta(\tau_1-\tau_2)\\
\displaystyle
\left\langle\eta(\tau_1)\eta(\tau_2)\cdots\eta(\tau_{2n})\right\rangle = 
\sum_{\pi}
\left\langle\eta_{\pi(1)}\eta_{\pi(2)}\right\rangle
\left\langle\eta_{\pi(3)}\eta_{\pi(4)}\right\rangle\cdots
\left\langle\eta_{\pi(2n-1)}\eta_{\pi(2n)}\right\rangle
\end{array}
\end{equation}
where the sum is over all permutations, $\pi$, of the index values
$1,2,3,\ldots,2n$. 

We shall assume that $U(x(\tau))$ is an ultra--local functional of $x(\tau)$,
i.e. doesn't contain any derivatives with respect to $\tau$ and discuss in our
conclusions what happens when $U(x(\tau))$ is, simply, a local functional of
$x(\tau)$. 

We are interested in computing the probability distribution of the limiting
value,  $x_\infty\equiv\lim_{\tau\to\infty} x(\tau)$. In this limit, the
Gaussian stochastic process, $\eta(\tau)$, is a Gaussian variable
$\eta_\infty\equiv\eta$ and  
the Langevin equation takes the form (to simplify notation we set
$x_\infty\equiv x$ and $\eta_\infty\equiv\eta$)
\begin{equation}
\label{equilibrium_langevin}
\eta = \frac{d U(x)}{dx}
\end{equation}
This relation indicates that $dU(x)/dx$ is drawn from a Gaussian
distribution. We are interested in the distribution, $\rho(x)$, of $x$ and
will try to determine it from its moments, $\langle x^l\rangle$. The
partition function is given by the expression
\begin{equation}
\label{partition_function}
Z = \int_{-\infty}^{\infty} dx d\eta\,e^{-\frac{\eta^2}{2}}\delta\left(\eta-\frac{dU}{dx}\right) = 
\int_{-\infty}^{\infty} dx\,\left|\frac{d^2 U}{dx^2}\right| e^{-\frac{1}{2}\left(\frac{dU}{dx}\right)^2}
\end{equation}
This holds only if $U''(x)\neq 0$ for all values of $x$. 
 If $U''(x)$ can vanish then the second
equality does not hold-a well known fact of freshman calculus: we must
partition the domain of integration into intervals, where the Jacobian is of
fixed sign. The point(s) where the Jacobian vanishes are ``excised'' and the
integrals computed by limiting procedure. We would like to explore in greater
detail the symmetries of the generating function $Z$, written as an integral
over $x$, when this limiting procedure is necessary. The reason is that this
toy model will be useful as a laboratory for field theories. In fact it
describes the boundary degree(s) of freedom of supersymmetric quantum
mechanics in one space dimension. 

The plan of our paper is the following:

In section~\ref{wp_SUSY} we show that the ``classical'' effective action,
obtained by exponentiating the Jacobian, is invariant under a transformation,
linear in $x$,whose parameter is an anticommuting variable, thereby 
displaying worldpoint supersymmetry. Since we must introduce a regularization
procedure, in order to integrate over the fermionic zeromodes, it is not obvious that this symmetry is a symmetry of the 
partition function, as well. We present an argument, under what circumstances
the regularized partition function, once the cutoff is removed, will, indeed, 
respect supersymmetry. 

To gain further insight, we study the  infinite set of relations between the 
moments, the zero--dimensional avatars of the Ward--Takahashi identities of 
field theory. 

This we do in section~\ref{WT_SUSY}. We first establish the identities that
are satisfied by the moments of the auxiliary field. These, however, depend
quite loosely on the particular dynamical model. The identities between the
moments of the ``physical variable'' are obtained by replacing the auxiliary
field in these identities by its equation of motion. 
We must, therefore, show that the moments of the physical variable exist, for these identities to be well-defined. 

 We compute these moments in two, different, ways: on the one hand we invert 
the ``Nicolai map'', $F=dU/dx$--which can be
done exactly, for the example we consider--on the other hand we compute the
moments of the scalar variable from the moments of $\exp(-(dU/dx)^2/2)$. 

In section~\ref{Num_SUSY} we
insert these moments in the identities satisfied by the auxiliary field and
deduce relations for the moments of the physical variable itself. These can be
evaluated either by using properties of special functions, or numerically.

In section~\ref{conclusions} we discuss the challenges that the generalization
to finite dimensions poses and how they might be tackled, as well as the
issues for describing ``target space'' supersymmetry in this formalism.  

\section{Worldpoint Supersymmetry}\label{wp_SUSY}
We would like to write 
\begin{equation}
\label{statistical_weight}
\left|\frac{d^2 U(x)}{dx^2}\right|
e^{-\frac{1}{2}\left(\frac{dU}{dx}\right)^2}\equiv e^{-S_\mathrm{eff}}
\end{equation}
One way, of course, would be to write 
$$
\left|\frac{d^2 U(x)}{dx^2}\right| = e^{\log\left|\frac{d^2
    U(x)}{dx^2}\right|} = e^{\frac{1}{2}\log\left(\frac{d^2 U(x)}{dx^2}\right)^2}
$$
In field theory this expression would become $\exp\left(\mathrm{Tr}\log
U''(x)\right)$ and, since $U(x)$ in field theory is a local functional of the
fields, this contribution to the action would be  non-local. 

On the other hand, as is well known, we may introduce it in a local way (in
field theory) if we use two anticommuting variables, $\psi_\alpha$,
$\alpha=1,2$:
\begin{equation}
\label{anticommuting_vars}
\left|\frac{d^2 U(x)}{dx^2}\right| = 
\int d\psi_1 d\psi_2 e^{\frac{1}{2}\psi_\alpha\varepsilon^{\alpha\beta}
U''(x)\psi_\beta}
\end{equation} 
with 
$$
\varepsilon^{\alpha\beta}\equiv\left(\begin{array}{cc} 0 & -1\\ 1 & 0\end{array}\right)
$$
The anticommuting variables introduced here are {\em not} ghosts, but as
``physical'' as $x$. There isn't any notion of spin so the spin--statistics
theorem is vacuous.(The only way they could be ghosts would be if $U''(x)$ 
were imaginary.)

The effective action, therefore, becomes
\begin{equation}
\label{Seff0}
S_\mathrm{eff}(x,\psi) = \frac{1}{2}\left(\frac{d U}{dx}\right)^2-\frac{1}{2}\psi_\alpha\varepsilon^{\alpha\beta}U''(x)\psi_\beta
\end{equation} 
We notice a mismatch: one commuting degree of freedom versus two anticommuting
degrees of freedom. We may restore equality by introducing another commuting
variable that we will call, for historical reasons, $F$~\cite{nicolai}:
\begin{equation}
\label{auxiliary_field}
e^{-\frac{1}{2}\left(\frac{dU}{dx}\right)^2} = 
\int_{-\infty}^{\infty} dF\,e^{-\frac{1}{2}F^2 + \mathrm{i}F\frac{dU}{dx}} = 
\mathrm{i}\int_{-\mathrm{i}\infty}^{\mathrm{i}\infty}
dF\,e^{\frac{1}{2}F^2-F\frac{dU}{dx}}
\end{equation}
Here we remark that the variable $F$, in the field theory generalization, has
an ultra--local propagator and that, were this propagator local, $F$ would be
a ghost. The final expression for the classical effective action, therefore,
is 
\begin{equation}
\label{Seff1}
S_\mathrm{eff}(x,\psi,F) = -\frac{1}{2}F^2 + F\frac{dU}{dx}
-\frac{1}{2}\psi_\alpha\varepsilon^{\alpha\beta}\frac{d^2U}{dx^2}\psi_\beta
\end{equation}
This action is invariant under the following transformation\cite{nicolai}:
\begin{equation}
\label{SUSY_transf}
\begin{array}{l}
\displaystyle
\delta x = \zeta_\alpha\varepsilon^{\alpha\beta}\psi_\beta\\
\displaystyle
\delta\psi_\alpha = \zeta_\alpha F\\
\displaystyle
\delta F = 0
\end{array}
\end{equation}
The parameter(s), $\zeta_\alpha$, $\alpha=1,2$, are anticommuting variables so
we remark that this transformation mixes commuting and anticommuting
variables. It is thus called a supersymmetry transformation. Since there is
one anticommuting pair of variables, $\zeta_\alpha$, this is ${\mathcal N}=1$
supersymmetry. 
It doesn't seem to depend on the explicit form of $U(x)$, which 
is the zero--dimensional avatar of  the
superpotential (this is due to the fact that we have introduced the auxiliary
field, $F$, which leads to a linear
realization~\cite{parisi_sourlas,nicolai}). 
When $\left\langle F\right\rangle\neq 0$, these relations imply that the
fermion becomes the zero--diensional avatar of the goldstino. 

If we compute the anticommutator, generated by the
transformations~(\ref{SUSY_transf}), we find that 
\begin{equation}
\label{SUSY_Q}
\left[\zeta Q,\eta Q\right] = 0\Leftrightarrow\left\{Q_\alpha,Q_\beta\right\}
= 0
\end{equation}
This relation implies that eigenstates of these operators with nonzero
eigenvalue are paired. It is silent about the existence, or not of eigenstates
of zero eigenvalue. If such latter states exist, then supersymmetry is
realized--if not, it is broken. The resolution depends on the dynamics. 

Since it is the dynamics that will interest us, 
we shall work with the partition function and the identities of its moments 
and not consider the algebra itself in what follows. 
The calculations are much more direct and we need make much fewer assumptions.

The ``equation of motion'' for $F$, obtained by varying the effective action,
is 
\begin{equation}
\label{eomF}
F = \frac{dU}{dx}
\end{equation} 
From eq.~(\ref{equilibrium_langevin}) we deduce that $F = \eta$, which means
that the ``auxiliary variable'', $F$, which was introduced to render the
supersymmetry transormations linear in $x$, is drawn from the same
distribution as the noise. 

If $U(x)$ is a quadratic function of $x$, then the relation between $F$ and
$x$ is linear. This is the case of the free theory (provided the relation is
invertible). If $U(x)$ is not
quadratic, the standard approach~\cite{nicolai,damgaard,dAFFV} is to
express the ``physical'' field, $x$, in terms of the ``auxiliary'' field, $F$
through a perturbation expansion about a reference
configuration. Supersymmetry implies identities between the correlation
functions that follow from Wick's theorem applied to the Gaussian distribution
of the auxiliary field. 

Here we would like to understand what happens to supersymmetry, in this
formalism,  if $\varepsilon^{\alpha\beta}U''(x)$ has zeromodes.

In the next
section we shall write down the stochastic identities for the case of the cubic
superpotential, without recourse to any perturbative expansion.
 
We shall use two, logically independent, methods: The first consists in
solving the equation of motion for the auxiliary field and expressing the
moments of the scalar in terms of the moments of the auxiliary field. 

In the second approach we shall compute the moments of the scalar from the
moments of the classical action of the scalar.
 We end up with identical expressions in both
cases--since we solved the equation of motion of the auxiliary field
exactly. This is just a test case. In more general situations we cannot solve
the equation for the auxiliary field and the second approach is more effective.

We conclude with a discussion of directions for further inquiry.

\section{Worldpoint Supersymmetry beyond the  classical action}\label{WT_SUSY}
An example that illustrates these issues is that of the zero--dimensional
model with cubic superpotential:
\begin{equation}
S = -\frac{1}{2}F^2+F\left(m^2 x + \frac{\lambda}{2}x^2\right)
-\frac{1}{2}\psi_\alpha\varepsilon^{\alpha\beta}\left(m^2 + \lambda x\right)\psi_{\beta}
\end{equation}
When we integrate out the ``fermions'' and the auxiliary variable, $F$, we
find
$$
Z = 
\int_{-\infty}^\infty dx\,
\underbrace{\mathrm{sign}\left(m^2+\lambda x\right)\left(m^2+\lambda
  x\right)}_{|m^2+\lambda x|}
 e^{-\frac{1}{2}\left(m^2 x+\frac{\lambda}{2}x^2\right)^2}
$$
(In fact here there's a sign issue that doesn't have anything to do with the
sign of the Jacobian, namely, when we integrate over two anticommuting
variables, $\psi_1$ and $\psi_2$
$$
\int d\psi_1 d\psi_2 e^{A \psi_1\psi_2} = 
\int d\psi_1 d\psi_2 (1+ A\psi_1\psi_2) = - A
$$
This is a global sign and doesn't play any role in the calculation of
correlation functions-we fix it once and for all.)
 
We remark that $U''(x)=m^2+\lambda x$ vanishes at $x=-m^2/\lambda$. 
If we  try to compute $\left\langle x^p\right\rangle$ by direct sampling:
\begin{equation}
\label{direct_sample}
\left\langle x^p\right\rangle = 
\frac{
\int dx\,x^p\mathrm{sign}\left(m^2+\lambda x\right)\left(m^2+\lambda x\right)
e^{-\frac{1}{2}\left(m^2x+\frac{\lambda}{2}x^2\right)^2}
}
{
\int dx\,\mathrm{sign}\left(m^2+\lambda x\right)\left(m^2+\lambda x\right)
e^{-\frac{1}{2}\left(m^2x+\frac{\lambda}{2}x^2\right)^2}
}
\end{equation}
we see that this expression isn't well-defined, since the na\"ive action 
\begin{equation}
\label{naive_action}
S_\mathrm{naive} =
\frac{1}{2}\left(m^2x+\frac{\lambda}{2}x^2\right)^2-\ln|m^2+\lambda x|
\end{equation}
is not bounded from below: $\ln|m^2+\lambda x|\to -\infty$ as $x\to
-m^2/\lambda$. We remark, further,  that the ``classical action'', 
$$
S\equiv \frac{1}{2}\left(m^2x+\frac{\lambda}{2}x^2\right)^2
$$
has a double well structure, with degenerate 
minima at $x=0$ and $x=-2m^2/\lambda$ and a maximum at 
$x=-m^2/\lambda$. At this maximum the (Majorana) ``mass'' of the
``fermion'' vanishes--``chiral symmetry'' might be realized. At the minima
``chiral symmetry'' is ``broken''--the ``fermion'' is ``massive'' and 
its mass is  equal to that of the ``scalar''--supersymmetry is unbroken. 
If we study the system in perturbation theory, we can only access the 
vicinity of the minima and we should find unbroken
``supersymmetry': $m_\mathrm{F}= U''(x^\ast)$ and 
$m_\mathrm{B}^2 =  (U''(x^\ast)^2 + U'(x^\ast)
U'''(x^\ast)) = m_\mathrm{F}^2$, since $x^\ast$ is root of
$U'(x^\ast)=0$;   broken ``chiral symmetry'' ($m_\mathrm{F}\neq 0$)
 and no hint of the instability at the maximum.   
(If we draw $e^{-S}$, we notice that it has support on the whole real axis.)
When we try to compute the moments beyond  perturbation theory, however,
we will generate configurations around the maximum--where the Jacobian 
vanishes. 
In this case we need to understand, whether the theory is irretrievably sick
beyond perturbation theory, or can be salvaged. We start to suspect that far
from sickness this is a sign of health: we have tried  to ``integrate out'' a
 ``massless particle''-it shouldn't be surprising
 that we run into trouble, since  we're moving along flat directions in 
field space. 

 These are the zero dimensional avatars of the zero modes
of the Dirac operator in field theory. So, following standard procedure, we
must omit the point $x=-m^2/\lambda$, when we evaluate the Jacobian
(i.e. when we evaluate the fermionic determinant) and ``integrate over the
zeromode''. Here this amounts to ``excising'' the point $x=-m^2/\lambda$,
i.e. evaluating the following partition function 
\begin{equation}
\label{Zepsilon}
\begin{array}{l}
\displaystyle
Z_{\varepsilon} = 
\int_{-\infty}^{-\frac{m^2}{\lambda}-\varepsilon} dx\,
\underbrace{\mathrm{sign}\left(m^2+\lambda x\right)}_{=-1}\left(m^2+\lambda x\right)
e^{-\frac{1}{2}\left(m^2x + \frac{\lambda}{2}x^2\right)^2} +\\
\displaystyle
\int_{-\frac{m^2}{\lambda}+\varepsilon}^\infty dx\,
\underbrace{\mathrm{sign}\left(m^2+\lambda x\right)}_{=+1}\left(m^2+\lambda x\right)
e^{-\frac{1}{2}\left(m^2x + \frac{\lambda}{2}x^2\right)^2}
\end{array}
\end{equation}
and using it to compute the moments $\langle x^p\rangle$. Since we are not
integrating over the entire real line--we have deleted the interval
$[-m^2/\lambda-\varepsilon,-m^2/\lambda + \varepsilon]$--we have broken
supersymmetry explicitly~\cite{stochastic_app,parisi_sourlas}. However we are
interested in the limit, $\varepsilon\to 0$ and would like to understand,
whether, in this limit, supersymmetry is recovered, or not. 

The answer, in fact, is negative--and it isn't hard to see why. The reason is
that the zero of the Jacobian is not, also, a zero of the action. Were this the case, then supersymmetry would be recovered. The proof of this statement goes as
follows: 

In eq.~(\ref{Zepsilon}) the Jacobian no
longer vanishes in each integral. Therefore we can perform the change of
variables, $F = m^2x + (\lambda/2)x^2$, in each and obtain the
expression
\begin{equation}
\label{Zepsilon_u}
Z_\varepsilon =
-\int_{\infty}^{-\frac{m^4}{2\lambda}+\frac{\varepsilon^2\lambda}{2}} dF\,
e^{-F^2/2} + 
\int_{-\frac{m^4}{2\lambda}+\frac{\varepsilon^2\lambda}{2}}^\infty dF\, e^{-F^2/2}= 
2\int_{-\frac{m^4}{2\lambda}+\frac{\varepsilon^2\lambda}{2}}^\infty dF\,
e^{-F^2/2} 
\end{equation}
This is the exact, regularized, partition function for the auxiliary field.

For $\varepsilon$ finite,  even if we take the limit
$\varepsilon\to 0$, with $m$ and $\lambda$ fixed, the 
lower limit of integration tends to a finite value, 
$-m^4/(2\lambda)$. This signals supersymmetry breaking, in the
limit $\varepsilon\to 0$.  The reason is that, as $\varepsilon\to 0$, the
curve $F= m^2 x + (\lambda/2) x^2$ stays fixed--and the minimum of the right
hand side is not equal to zero. 

This, however, is easily arranged. It suffices to perform the following change
of variables:
\begin{equation}
\label{surface_term}
F = c + m^2 x +(\lambda/2)x^2 = c-\frac{m^4}{2\lambda}+\frac{\lambda}{2}\left(x+\frac{m^2}{\lambda}\right)^2
\end{equation}
and fix the constant $c$ by the condition that, in the limit $\varepsilon\to
0$, the lower bound of the range of integration is at zero. In this case we
immediately find that $c=m^4/(2\lambda)$. For this value of the coefficient of
the linear term, $F$ becomes a
perfect square, whose (double) zero coincides with the (simple) zero of the
Jacobian. 

It would seem that supersymmetry has been thereby restored. Alas, this is not
the case: having rendered the auxiliary field a perfect square means that its
range is not the whole real axis, but, only, the non-negative part of it. 
Therefore the integration domain has a boundary--and that leads to supersymmetry
breaking:
\begin{equation}
\label{vevFu}
\left\langle F\right\rangle = \frac{2\int_0^\infty dF\,
  e^{-F^2/2}F}{2\int_0^\infty dF\,e^{-F^2/2}}\neq 0
\end{equation}
We realize that the fundamental reason is that the curve, $F(x)$,
eq.~(\ref{surface_term}), has a global maximum--or minimum, depending on the
sign of $\lambda$--that prevents us from extending the integration range to
the full real axis. We would need it to have, at most, an inflection point. 
An example of this is provided by the quartic superpotential, 
\begin{equation}
\label{quartic}
U(x) = \frac{m^2}{2}x^2 + \frac{\lambda}{4!}x^4 + cx\Leftrightarrow F =
\frac{dU}{dx} = m^2 x + \frac{\lambda}{6}x^3 + c
\end{equation}
If $m^2>0$, $\lambda>0$ then the Jacobian, $U''(x)=m^2+(\lambda/2)x^2$ is
positive definite, therefore the partition function becomes a Gaussian
integral, centered at zero and whose integration range is the whole real
axis. Supersymmetry is realized in the Wigner mode. 

If $m^2<0, \lambda>0$, then the Jacobian vanishes at two points,
$x=\pm\sqrt{-2m^2/\lambda}$. Its sign is positive outside of the interval
$(-\sqrt{-2m^2/\lambda},\sqrt{-2m^2/\lambda})$ and negative within. Therefore,
the partition function takes the following form
\begin{equation}
\label{quarticpf}
\begin{array}{l}
\displaystyle
Z_\varepsilon = \int_{-\infty}^{-\sqrt{-2m^2/\lambda}-\varepsilon}
  dx\,\left(m^2 + \frac{\lambda}{2}x^2\right) e^{-\frac{1}{2}F^2} -
 \int_{-\sqrt{-2m^2/\lambda}+\varepsilon}^{\sqrt{-2m^2/\lambda}-\varepsilon}
  dx\,\left(m^2 + \frac{\lambda}{2}x^2\right) e^{-\frac{1}{2}F^2}  + \\
\displaystyle
\hskip1.3truecm
\int_{\sqrt{-2m^2/\lambda}+\varepsilon}^\infty
dx\,\left(m^2+\frac{\lambda}{2}x^2\right) e^{-\frac{1}{2}F^2}
\end{array}
\end{equation}
We may perform the change of variables, $F=c+m^2 x + (\lambda/6)x^3$, in each
integral. The partition function becomes 
\begin{equation}
\label{quarticpf1}
\begin{array}{l}
\displaystyle
Z_\varepsilon = \int_{-\infty}^{F(-\sqrt{-2m^2/\lambda}-\varepsilon)}
dF\,e^{-F^2/2}
-\int_{F(-\sqrt{-2m^2/\lambda}+\varepsilon)}^{F(\sqrt{-2m^2/\lambda}-\varepsilon)}
dF\,e^{-F^2/2} + \\
\displaystyle
\hskip1.3truecm
\int_{F(\sqrt{-2m^2/\lambda}+\varepsilon)}^\infty dF\,e^{-F^2/2}
\end{array}
\end{equation}
and we notice that, since $F(x)$ is decreasing in the middle integral,
$F(-\sqrt{-2m^2/\lambda})>F(\sqrt{-2m^2/\lambda})$. Therefore, in the limit
$\varepsilon\to 0$, which is smooth, the contribution of the middle intgral
cancels out the contribution from the overlap between the two other integrals,
we are left with an integral over the whole real axis--and supersymmetry
 is restored. The linear term doesn't play any role at all, as far as
 realization or not of supersymmetry is concerned--it can only affect how
 supersymmetry is broken: 
For the cubic superpotential, even if its coefficient doesn't take the special 
value, $m^4/(2\lambda)$, 
the range of integration for $F$ still cannot cover the whole real axis and
supersymmetry will be broken. For the quartic superpotential the linear term
is completely invisible, as far as supersymmetry breaking is concerned. 

This analysis settles the issue of supersymmetry breaking, in principle, for
these models. We would like, however, to understand the consequences for the
moments of the scalar.

In the remainder of this section we shall focus on the cubic superpotential
and establish how the relations between the moments of the
auxiliary field, $F$, even when supersymmetry is broken in the way we have
seen, imply, nonetheless, relations between the moments of the scalar. 

We start from the partition function of the auxiliary field itself, 
since it is now in a well-defined
form and deduce recursion relations between its moments. These are simple
enough, that we may solve them exactly. However they are relations between
moments of the auxiliary field--to render them effective, we need the moments
of the scalar.

In the following subsection we establish well-defined expressions for the
moments of the scalar variable, $x$, itself and use an exact inversion of the
``Nicolai map'', $F = dU/dx$, to express the moments of $x$ in terms of the
moments of $F$. The identities for the moments of the auxiliary field thus
become identities for the moments of the scalar. However these identities do
not carry additional information, since they simply express the Nicolai map,
which is exact. 

Such information is obtained from our calculation in the next
subsection~\ref{ScalMoms}. There we compute the moments of the scalar, without
recourse to the auxiliary field at all. We obtain in this way expressions that
involve moments with respect to an, {\em a priori}, completely different
distribution. Nevertheless, the way we conducted the transformations implies
that the moments thus obtained must, when substituted in the expression for
the auxiliary field, lead to the identities that it satisfies. In our example
we can see explicitly that we obtain the same expresion as that from inverting
the Nicolai map. In more complicated cases, however, this, second, approach is
the only one available. The simplest example is that of the quartic
superpotential. We check our expression for the moments by substituting them
in the identities satisfied by the auxiliary field. It is here that we really
needed to compute the integrals numerically. We compute several identities,
namely $\left\langle F\right\rangle =0$ while varying the couplings and verify
that supersymmetry is, indeed, realized, to numerical accuracy. 

The true payoff, of course, lies in the moments for the scalar, for whose 
distribution we do not have an explicit expression, but which we can, in
principle, reconstruct. We can compute, in particular, for the cubic
superpotential, the exact expression
for the connected fourth moment, (that, in the field theory context, 
would control the scattering of two scalars to two scalars) and show that, for
the cubic superpotential, it 
never vanishes. Therefore the distribution for the scalar is always 
non-Gaussian.

\subsection{The moments of the auxiliary field}\label{auxfieldmoms}
From the expression of the partition function for the auxiliary field,
eq.~(\ref{Zepsilon_u}) 
 we deduce the following recursion relation for the
moments
\begin{equation}
\label{SDEq}
\begin{array}{l}
\displaystyle
\frac{1}{Z_\varepsilon}2\int_{c-\frac{m^4}{2\lambda}+\frac{\varepsilon^2\lambda}{2}}^\infty
dF\frac{d}{dF}\left(F^k e^{-F^2/2}\right) =
\underbrace{-\left(c-\frac{m^4}{2\lambda}+\frac{\varepsilon^2\lambda}{2}\right)^k\frac{2}{Z_\varepsilon}e^{-\frac{1}{2}\left(c-\frac{m^4}{2\lambda}+\frac{\varepsilon^2\lambda}{2}\right)^2}}_{B_k}\\
\displaystyle
\frac{1}{Z_\varepsilon}2\int_{c-\frac{m^4}{2\lambda}+\frac{\varepsilon^2\lambda}{2}}^\infty
dF\frac{d}{dF}\left(F^k e^{-F^2/2}\right) =
k\left\langle F^{k-1}\right\rangle_F - \left\langle F^{k+1}\right\rangle_F = B_k
\end{array}
\end{equation}
whence we deduce that the odd moments and the even moments decouple:
\begin{equation}
\label{recmoms}
\begin{array}{l}
\displaystyle
\left\langle F^{2l+1}\right\rangle_F = 2l\left\langle F^{2l-1}\right\rangle_F
-B_{2l}\\
\displaystyle
\left\langle F^{2l}\right\rangle_F = (2l-1)\left\langle F^{2(l-1)}\right\rangle_F 
-B_{2l-1}\\
\end{array}
\end{equation}
Since $Z_\varepsilon$ exists, $\left\langle 1\right\rangle_F=1$, $\left\langle
F\right\rangle_F$ and $\left\langle F^2\right\rangle_F$ exist as well, 
all higher moments of the auxiliary field are uniquely specified by the 
first and second moments. 

\begin{figure}[thp]
\begin{center}
\subfigure{\includegraphics[scale=0.7]{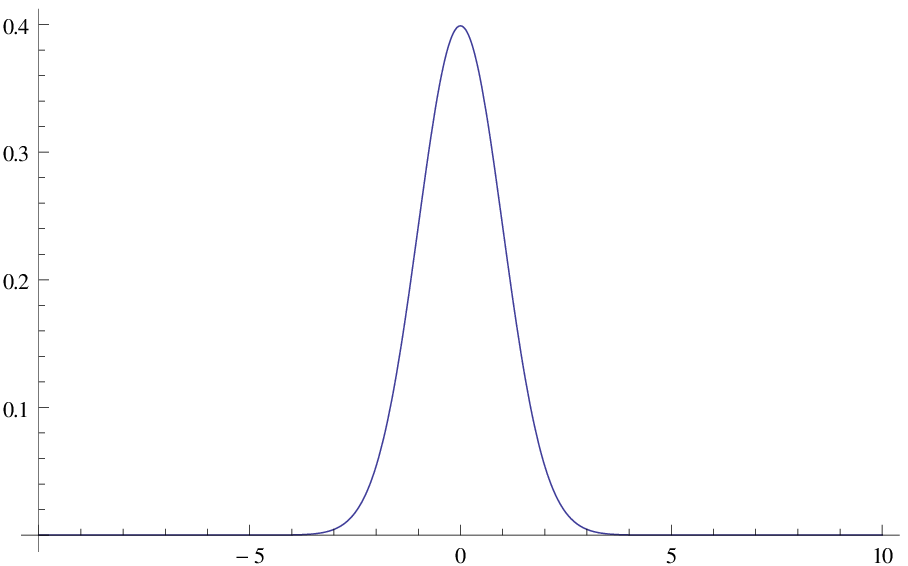}}
\subfigure{\includegraphics[scale=0.7]{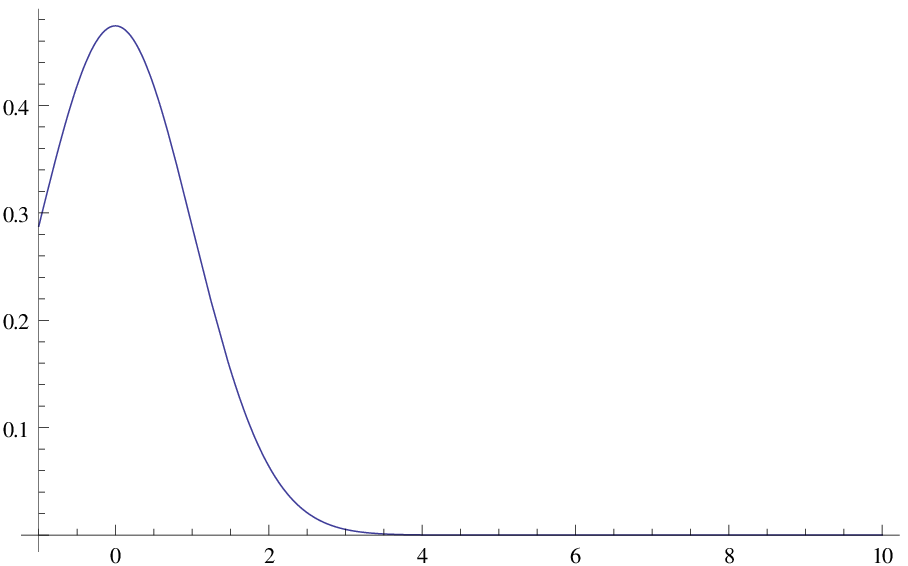} }
\subfigure{\includegraphics[scale=0.7]{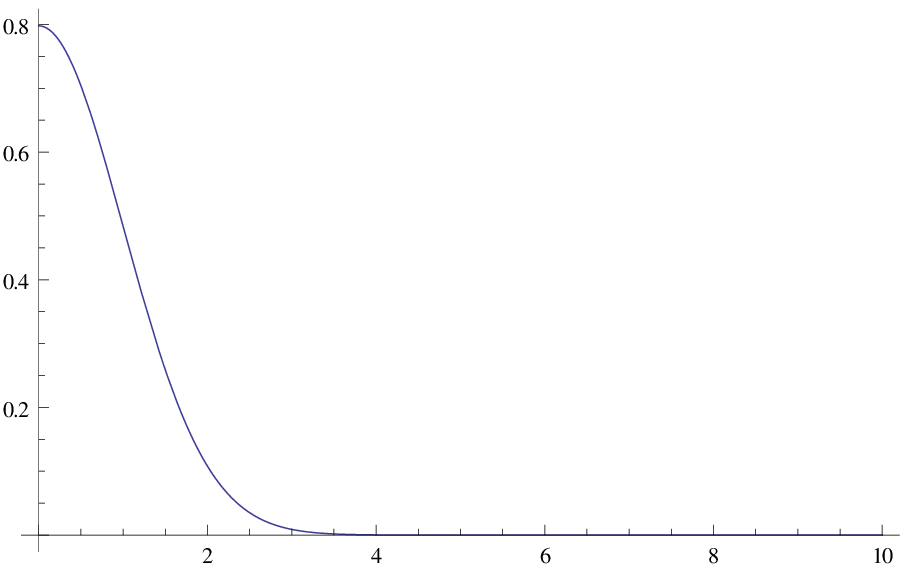}}
\subfigure{\includegraphics[scale=0.7]{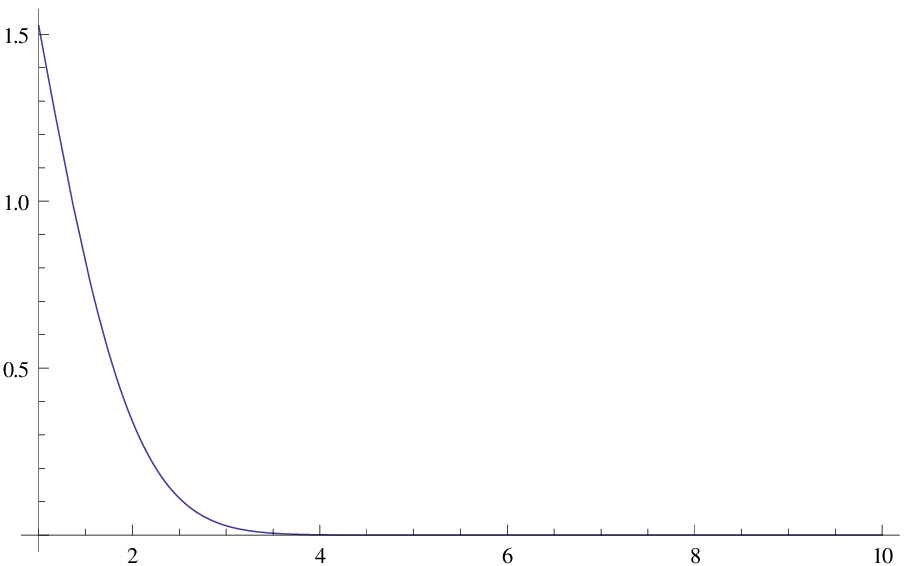} }
\end{center}
\caption[]{ Some examples for the distribution of the auxiliary field,
  $\rho(F)$, vs. $F\geq (c-(m^4/(2\lambda))$, 
for different values of $c-(m^4/(2\lambda))$. 
(a) $c-m^4/(2\lambda)=-10$--here $F^\ast=0\approx\left\langle F\right\rangle$; 
(b) $c-(m^4/(2\lambda))=-1$--here $F^\ast=0\neq\left\langle F\right\rangle$, 
$F^\ast$ is the typical value. 
(c) $c=m^4/(2\lambda)$--here, too, $F^\ast=0\neq\left\langle F\right\rangle$.  
(d) $c-(m^4/(2\lambda))=1$--here $F^\ast=0$ doesn't lie in the sampling
domain, $F^\ast\neq\left\langle F\right\rangle\neq 0$.} 
\label{auxFdist}
\end{figure}

\begin{figure}[thp]
\begin{center}
\includegraphics[scale=0.8]{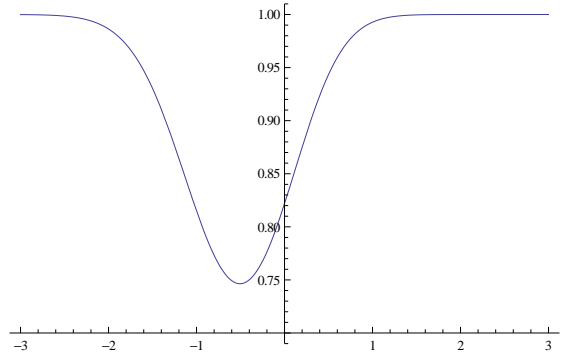}
\end{center}
\caption[]{$\rho(\left\langle F\right\rangle)/\rho(F^\ast)$ as a function of 
$c-(m^4/(2\lambda))$.}
\label{typicalratio}
\end{figure}

These identities do not constrain the 1--point function, $\left\langle
F\right\rangle_F$, whose expression is 
\begin{equation}
\label{1pointF}
\left\langle F\right\rangle_F = \frac{1}{Z_{\varepsilon\to
    0}}
2\int_{c-\frac{m^4}{2\lambda}}^\infty 
dF\,e^{-F^2/2}F = 
\frac{
e^{-\frac{1}{2}\left(c-\frac{m^4}{2\lambda}\right)^2}}
{2\int_{c-\frac{m^4}{2\lambda}}^\infty dF\,e^{-F^2/2}}
\end{equation}
This result is interesting for several reasons: (a) A consequence of the
``classical equation of motion'', $F = U'(x)$, for the auxiliary field, was
that $F=\eta$, thus $\left\langle F\right\rangle = 0$. 
 This equation is, obviously, 
in contradiction with eq.~(\ref{1pointF}). The
resolution is that the symmetry is spontaneously  broken by the interaction
with the noise (parametrized by the fermions).  
 (b) If we draw the distribution, 
\begin{equation}
\label{rhoF}
\rho(F) = \frac{e^{-F^2/2}}{\int_{c-\frac{m^4}{2\lambda}}^\infty dF\,e^{-F^2/2}}
\end{equation}
for different values of the lower limit, $c-(m^4/(2\lambda)$,
cf. fig.~\ref{auxFdist}, 
we notice that its maximum is at
$F^\ast=0$, which is different from $\left\langle
F\right\rangle_F$. Thus, $\rho(\left\langle
F\right\rangle)<\rho(F^\ast)$. This means that, even though it is
$\left\langle F\right\rangle$ that controls, whether supersymmetry is broken,
or not, $F^\ast=0$ is the ``typical'' value that will be drawn from
$\rho(F)$. The ratio, 
\begin{equation}
\label{typical}
\frac{\rho(\left\langle F\right\rangle)}{\rho(F^\ast)} = 
\exp\left[
-\frac{1}{2}
\left(
\frac{e^{-\frac{1}{2}\left(c-\frac{m^4}{2\lambda}\right)^2}}
     {2\int_{c-\frac{m^4}{2\lambda}}^\infty dF\,e^{-F^2/2} }
\right)^2
\right]
\end{equation}
We plot this ratio, as a function of the control parameter,
$c-(m^4/(2\lambda))$,in fig.~\ref{typicalratio}.
The result is that, for $c-m^4/(2\lambda)\approx -0.3$, the ratio falls to $\sim 0.75$. For 
$|c-(m^4/(2\lambda))|>1$, this ratio is, practically, equal to one. But, for
these values of the control parameter, $\rho(F)$ is, also, very small, so 
large samples  are required. 

Therefore, if $c=m^4/(2\lambda)$, in which case $B_k=0$,
we would be tempted to conclude that supersymmetry is realized, even though it
is spontaneously  broken, if we couldn't distinguish the typical from the
average value. On the other hand, this might, also, imply that supersymmetry
breaking effects are much harder to detect than expected, due to this
background.  
This issue certainly deserves further
study in more realistic cases. In disordered systems this was noted in
ref.~\cite{derrida_hilhorst}; in ref.~\cite{crisanti_etal} it was shown that
it is, indeed, possible to distinguish the typical from the average value of
certain correlators in disordered spin chains. 
If $c\neq m^4/(2\lambda)$, then $B_k\neq 0$ and the identities satsfied by the
moments of the auxiliary field acquire ``anomalous'' terms. 
Nevertheless, from the relation between the auxiliary field and the scalar,
eqs.~(\ref{recmoms}) lead to constraints on the moments of the scalar.

We shall present two approaches 
 for computing its moments, $\left\langle x^p\right\rangle$, none of which
 entails an expansion in any coupling constant, but is exact, (to machine precision). 
  The first simply relies on the fact
that we can solve quadratic algebraic equations explicitly; the second dispenses
with this requirement, but relies on the ability of evaluating 
integrals numerically.

\subsection{Inverting the Nicolai map}\label{NicmapInv}
The relations obtained above are quite model independent--the existence of the
boundary limit of integration, $B$, doesn't depend on a specific model, only
on a class as a whole--to give them content we
must replace the auxiliary field by the derivative of the superpotential under
study and define how we compute the moments of the scalar. 

 In this subsection we shall assume that we can solve the equation of
motion of the auxiliary field, 
$$
F = \frac{dU}{dx}
$$
exactly. This is what is meant by ``inverting the Nicolai map''. We shall use
this knowledge to write the regularized 
moments, $\left\langle x^p\right\rangle_\varepsilon$ of the scalar in a form
that is particularly suited towards numerical calculation and shall try to
check, whether these moments, when inserted in the expression $\left\langle
(dU/dx)^q\right\rangle_\varepsilon$, in the limit $\varepsilon\to 0$, lead to
identities consistent with Wick's theorem.

Since we are particularly interested in the role of the 
linear term of the superpotential,
we shall include it, with an arbitrary coefficient. 

The expression for the regularized moments of the
scalar variable, $x$ reads as follows:
\begin{equation}
\label{regmoms}
\left\langle x^p\right\rangle_\varepsilon = 
\frac{ 
-\int_{-\infty}^{-\frac{m^2}{\lambda}-\varepsilon}dx\,(m^2+\lambda x)x^p
e^{-\frac{1}{2}\left(c+m^2 x + \frac{\lambda}{2}x^2\right)^2} + 
\int_{-\frac{m^2}{\lambda}+\varepsilon}^\infty dx\,
(m^2+\lambda x) x^p e^{-\frac{1}{2}\left(c+m^2 x +
  \frac{\lambda}{2}x^2\right)^2}
}
{
-\int_{-\infty}^{-\frac{m^2}{\lambda}-\varepsilon}dx\,(m^2+\lambda x)
e^{-\frac{1}{2}\left(c+m^2 x + \frac{\lambda}{2}x^2\right)^2} + 
\int_{-\frac{m^2}{\lambda}+\varepsilon}^\infty dx\,
(m^2+\lambda x) e^{-\frac{1}{2}\left(c+m^2 x + \frac{\lambda}{2}x^2\right)^2}
}
\end{equation} 
We perform the change of variables, 
$$
F = c + m^2 x + \frac{\lambda}{2}x^2\Leftrightarrow dF = dx (m^2 + \lambda x)
$$
whereupon the moments are computed as
\begin{equation}
\label{regmoms1}
\left\langle x^p\right\rangle_\varepsilon = 
\frac{\int_{c-\frac{m^4}{2\lambda}+\frac{\varepsilon^2\lambda}{2}}^\infty
dF e^{-F^2/2}\left[x_-(F)^p + x_+(F)^p\right]}{
2\int_{c-\frac{m^4}{2\lambda}+\frac{\varepsilon^2\lambda}{2}}^\infty
dF e^{-F^2/2}}
\end{equation}
where 
\begin{equation}
\label{Nicolai_map_roots}
x_\pm(F) = -\frac{m^2}{\lambda}\pm\sqrt{\frac{2}{\lambda}\left[F-\left(c-\frac{m^4}{2\lambda}\right)\right]}
\end{equation}
We end up with the following compact expression for the regularized moments,
$\left\langle x^p\right\rangle_\varepsilon$:
\begin{equation}
\label{sum_roots}
\begin{array}{l}
\displaystyle
\left\langle x^p\right\rangle_\varepsilon = 
\left(-\frac{m^2}{\lambda}\right)^p 2\sum_{k=0,k\,\mathrm{even}}^p\left(\begin{array}{c}p\\ k\end{array}\right)
\frac{
\int_{c-\frac{m^4}{2\lambda}+\frac{\varepsilon^2\lambda}{2}}^\infty
dF
e^{-F^2/2}\left(\frac{2\lambda}{m^4}\left(F-c+\frac{m^4}{2\lambda}\right)\right)^{k/2}
}{
2\int_{c-\frac{m^4}{2\lambda}+\frac{\varepsilon^2\lambda}{2}}^\infty
dF
e^{-F^2/2}
} = \\
\displaystyle
\hskip1.6truecm
\left(\frac{m^2}{\lambda}\right)^p(-)^p\sum_{k=0}^p
\left(\begin{array}{c}
  p\\ 2k\end{array}\right)\left(\frac{2\lambda}{m^4}\right)^k\left\langle
  F^k\right\rangle_F
\end{array}
\end{equation}
where $\left\langle F^k\right\rangle_F$ are computed using the
distribution~(\ref{rhoF}). 

From these expressions we can, in principle, recover the density, $\rho(x)$
and the action, $S(x)$. 

It must be stressed that $\left\langle x^l\right\rangle$ are the {\em exact}
moments. They are the coefficients in the Taylor series expansion of the
Fourier transform, $\widetilde{\rho}(k)$, of the density, 
$$
\rho(x) = \frac{e^{-S(x)}}{\int_{-\infty}^\infty du\,e^{-S(u)}}
$$
around
$k=0$:
\begin{equation}
\label{moments_density}
\begin{array}{l}
\displaystyle
\widetilde{\rho}(k) = \int_{-\infty}^\infty
dx\,e^{\mathrm{i}kx}\,\rho(x)\Leftrightarrow
\mathrm{i}^{-l}\left.\frac{d^l\widetilde{\rho}(k)}{dk^l}\right|_{k=0} =
\int_{-\infty}^\infty dx\,x^l\,\rho(x)=\left\langle x^l\right\rangle\Rightarrow\\
\displaystyle
\widetilde{\rho}(k)=\sum_{l=0}^\infty\frac{\left(\mathrm{i}k\right)^l}{l!}\left\langle
x^l\right\rangle\Rightarrow
\rho(x)=\int_{-\infty}^\infty\frac{dk}{2\pi}\,e^{-\mathrm{i}kx}\,\widetilde{\rho}(k)
\end{array}
\end{equation}
While the (exact) moments in eq.~(\ref{sum_roots}) 
 are certainly well defined quantities, whether they do, indeed, define a
 density, $\rho(x)$ and, thus, an ``action'', $S(x)=-\ln\rho(x)$, for all
 values of the ``coupling'', $2\lambda/|m^2|^2$, is a question that deserves
 further study, that will be reported elsewhere. 
 
In these calculations we relied on the fact that we could solve $F=dU/dx$
exactly. 
There are, however, situations when we cannot invert the Nicolai map as
efficiently as done in this case. 
An example would be that of a quartic superpotential or 
many superfields. Then we would have to solve a cubic equation, or a 
system of non-linear equations-and this cannot be
done as easily as for one quadratic equation. So we need to develop methods for
computing the moments of the scalar that do not rely on the Nicolai map.

To test the approach we shall apply it to the case at hand, where a 
direct comparison is possible and to the quartic superpotential.

\subsection{A new measure for the scalar's moments}\label{ScalMoms}
Our calculations were based on inverting the ``Nicolai map'',
$u=dU/dx$--however this inversion didn't involve any approximations at all,
since we were ``lucky'' that we had to solve a quadratic equation, whose roots
are known analytically. 
Had we  chosen not to invert it--or could not invert it exactly--we would have
to  compute the moments, $\left\langle x^p\right\rangle$ and use them 
to compute the moments, $\left\langle F^q\right\rangle$, of the auxiliary
field. Of course we could have computed the moments in a 
perturbative expansion~\cite{nicolai,dAFFV}, but this would weaken
considerably the conclusions we could reach. With numerical methods available
we can explore what happens beyond perturbation theory. 

It is instructive to compare the two approaches in this case. 

The idea is to compute the moments $\left\langle x^p\right\rangle$ in terms of
the moments of 
\begin{equation}
\label{new_weight_momx}
\rho(y)=\exp\left(-\frac{1}{2}
\left(
c-\frac{m^4}{2\lambda}+\frac{\lambda}{2}y^2\right)^2\right)\equiv e^{-S(y^2)}
\end{equation} 
Our starting point is eq.~(\ref{regmoms}). We set 
$$
y\equiv x +\frac{m^2}{\lambda}
$$
and write the numerator as follows:
\begin{equation}
\label{new_num_moms}
\lambda\left(
-\int_{-\infty}^{-\varepsilon}dy\,e^{-S(y^2)}
y\left(y-\frac{m^2}{\lambda}\right)^p
  +
 \int_{\varepsilon}^{\infty}dy\,e^{-S(y^2)}y\left(y-\frac{m^2}{\lambda}\right)^p\right)
\end{equation}
The denominator is obtained setting $p=0$ in this expression. Once more, the
$\varepsilon\to 0$ limit is smooth, so the moments 
$\left\langle x^p\right\rangle$ may be computed by 
\begin{equation}
\label{new_vev_x}
\begin{array}{l}
\displaystyle
\lim_{\varepsilon\to 0}\left\langle x^p\right\rangle_\varepsilon = 
\lim_{\varepsilon\to 0}\frac{
-\int_{-\infty}^{-\varepsilon}dy\,e^{-S(y^2)}y\left(y-\frac{m^2}{\lambda}\right)^p
  +
 \int_{\varepsilon}^{\infty}dy\,e^{-S(y^2)}y\left(y-\frac{m^2}{\lambda}\right)^p
}
{
 \int_{\varepsilon^2}^{\infty}dy\,e^{-S(y)}
} =\\
\displaystyle
\hskip1.6truecm
\frac{ 
\int_0^\infty
dy\,e^{-\frac{1}{2}\left(c-\frac{m^4}{2\lambda}+\frac{\lambda}{2}y^2\right)^2}
y\left[(-)^p\left(y+\frac{m^2}{\lambda}\right)^p + \left(y-\frac{m^2}{\lambda}\right)^p\right]
}
{\int_0^\infty dy\,e^{-\frac{1}{2}\left(c-\frac{m^4}{2\lambda}+\frac{\lambda}{2}y\right)^2}
}
\end{array}
\end{equation}
The numerator can be simplified:
\begin{equation}
\label{new_vev_x_num}
\begin{array}{l}
\displaystyle
\left(-y-\frac{m^2}{\lambda}\right)^p + \left(y-\frac{m^2}{\lambda}\right)^p = 
\sum_{k=0}^p\left(\begin{array}{c}p\\k\end{array}\right)
\left(-\frac{m^2}{\lambda}\right)^{p-k}
\left[(-y)^k+y^k\right] =\\
\displaystyle
\hskip3truecm
2\left(-\frac{m^2}{\lambda}\right)^p\sum_{k=0}^p\left(\begin{array}{c}p\\ 2k\end{array}\right)\left(\frac{\lambda}{m^2}\right)^{2k}y^{2k}
\end{array} 
\end{equation}
This leads to the change of variables $2y dy=d(y^2)$ in the
numerator. After some further straightforward algebra, we end up with the
expression 
\begin{equation}
\label{new_vev_x_moms}
\left\langle x^p\right\rangle = 
\left(\frac{m^2}{\lambda}\right)^p(-)^p
\sum_{k=0}^p\left(\begin{array}{c}p\\ 2k\end{array}\right)\left(\frac{2\lambda}{m^4}\right)^k
\frac{\int_0^\infty
  dy\,e^{-\frac{1}{2}\left(c-\frac{m^4}{2\lambda}+y\right)^2}y^k}
{
\int_0^\infty dy\,e^{-\frac{1}{2}\left(c-\frac{m^4}{2\lambda}+y\right)^2}}
\end{equation}
which, we perceive, after the fact, is identical to eq.~(\ref{sum_roots})! 
We didn't need to solve the equation $F=dU/dx$--we only used the knowledge of 
the root, $x=-m^2/\lambda$, of $d^2 U/dx^2=m^2+\lambda x=0$, around which we 
needed to impose the excision. 

For the quartic superpotential, with $m^2<0,\lambda>0$, the procedure is the following: We write 
$U''(x)=m^2 + (\lambda/2)x^2 = (\lambda/2)(x-x_1)(x-x_2)$, with $x_1<x_2$. If
we define the ``reduced partition function'',$\zeta_{a,b}$, by the expression
\begin{equation}
\label{zeta_ab}
\zeta_{a,b}\equiv\int_a^b dx\,e^{-\frac{1}{2}(dU/dx)^2}
\end{equation}
the moments, $\left\langle x^p\right\rangle$, are given by the
expression
\begin{equation}
\label{vev_x_moms_quartic}
\begin{array}{l}
\displaystyle
\left\langle x^p\right\rangle = 
\frac{\left\langle U'' x^p\right\rangle_{-\infty,\infty}}
     { \left\langle U''\right\rangle_{-\infty,\infty} } \equiv\\
\displaystyle
\lim_{\varepsilon\to 0}\left\langle x^p\right\rangle_\varepsilon = 
\lim_{\varepsilon\to 0}
\frac{ 
\zeta_{-\infty,x_1-\varepsilon}\left\langle U''
x^p\right\rangle_{-\infty,x_1-\varepsilon} -
\zeta_{x_1+\varepsilon,x_2-\varepsilon}\left\langle U''
x^p\right\rangle_{x_1+\varepsilon,x_2-\varepsilon} +
\zeta_{x_2+\varepsilon,\infty}\left\langle U''x^p\right\rangle_{x_2+\varepsilon,\infty}
}
{ 
\zeta_{-\infty,x_1-\varepsilon}\left\langle U''
\right\rangle_{-\infty,x_1-\varepsilon} -
\zeta_{x_1+\varepsilon,x_2-\varepsilon}\left\langle U''
\right\rangle_{x_1+\varepsilon,x_2-\varepsilon} +
\zeta_{x_2+\varepsilon,\infty}\left\langle U''\right\rangle_{x_2+\varepsilon,\infty}
}
\end{array}
\end{equation}
where $\left\langle{\mathcal O}\right\rangle_{a,b}$ denotes average with respect
  to $\zeta_{a,b}$. Of course here this is just a convenient shorthand--it
  is more efficient to use Simpson's rule than Monte Carlo (we've used
  both). But for the more general cases, these expressions generalize
  better. They are ``templates'': for concrete applications it is useful to
  try to simplify them, in order to generate efficient code, as was, already,
  done,  for the cubic superpotential. 
 
 In the following section we shall study the identities that these moments 
satisfy.

\section{Moments and Identities}\label{Num_SUSY}
Let us start with the cubic superpotential.

The expressions for the moments display, first of all, scaling: 
\begin{equation}
\label{scaling}
\left\langle x^p\right\rangle_\varepsilon = \left(\frac{m^2}{\lambda}\right)^p f_p\left(\frac{\lambda}{m^4}\right)
\end{equation}
The prefactor,
$(m^2/\lambda)^p$ indicates 
  the ``canonical dimension'' of the $p-$th moment--which 
coincides with its ``scaling dimension''. The ``scaling function'',
$f_p(\cdot)$, is a function only of the combination, $\lambda/m^4$. The
combination $m^2/\lambda$ simply sets the scale, it is the combination
$\lambda/m^4$ that plays the role of the coupling constant. In other words, the
combinations that are relevant for describing the moments are not $m^2$ and
$\lambda$, but $m^2/\lambda$ and $\lambda/m^4$. 

The moments are perfectly regular functions of the cutoff, $\varepsilon$, and
have a smooth limit, as it is removed, $\varepsilon\to 0$. The coefficient,
$c$, of the linear term of the superpotential simply affects the value of
integration limit--even if this can be pushed to infinity in the denominator
(thanks to the factor of 2), it survives in the numerator. However, if 
\begin{equation}
\label{spont_susy_b}
c=\frac{m^4}{2\lambda}
\end{equation}
then the identities satisfied by the auxiliary field, $F$ imply that
supersymmetry is spontaneously broken: $\left\langle F\right\rangle_F\neq 0$
but all other identities that express the Gaussian distribution for $F$ are
exactly satisfied. 

If $c\neq m^4/(2\lambda)$, then the fermion is still the goldstino, but the WT
identities have anomalies. 

If we compute $\left\langle F\right\rangle$, in the limit $\varepsilon\to 0$ and for $c=m^4/(2\lambda)$,
we find that 
\begin{equation}
\label{vevF1}
\left\langle F\right\rangle = \frac{m^4}{2\lambda} + 
m^2\left\langle x\right\rangle +
\frac{\lambda}{2}\left\langle x^2\right\rangle = 
\underbrace{\frac{m^4}{2\lambda}-\frac{m^4}{\lambda}+\frac{m^4}{2\lambda}
}_{=0} + 
\underbrace{\frac{\int_0^\infty dF\, e^{-F^2/2} F}{2\int_0^\infty dF\,e^{-F^2/2}}
}_{\neq 0}
\end{equation}
identical with the expression~(\ref{1pointF}), when $c=m^4/(2\lambda)$. 

Similarly, if we compute $\left\langle F^2\right\rangle$,when
$c=m^4/(2\lambda)$, in the limit $\varepsilon\to 0$, we find
\begin{equation}
\label{vevF2}
\begin{array}{l}
\displaystyle
\left\langle F^2\right\rangle =
\left\langle\frac{\lambda^2}{4}\left(x+\frac{m^2}{\lambda}\right)^4\right\rangle
= \\
\displaystyle
\frac{\lambda^2}{4}\left(
\left\langle x^4\right\rangle + 
4\left\langle x^3\right\rangle\frac{m^2}{\lambda} + 
6\left\langle x^2\right\rangle\left(\frac{m^2}{\lambda}\right)^2 + 
4\left\langle x\right\rangle\left(\frac{m^2}{\lambda}\right)^3 +
\left(\frac{m^2}{\lambda}\right)^4\right)  
\end{array}
\end{equation}
Substituting the expressions for the moments~(\ref{sum_roots}), we find,
indeed, that $\left\langle F^2\right\rangle = \left\langle F^2\right\rangle$,
i.e. that the right hand side of eq.~(\ref{vevF2}) is equal to 1. 

These are but two examples of eq.~(\ref{SDEq}), which implies that
the moments, $\left\langle x^p\right\rangle$, 
 of the scalar, given by eq.~(\ref{sum_roots}) or~(\ref{new_vev_x}),
 satisfy the following identities
\begin{equation}
\label{master_identity_cubic}
\left\langle\left(c-\frac{m^4}{2\lambda}+\frac{\lambda}{2}\left(x+\frac{m^2}{\lambda}\right)^2\right)^{k+1}\right\rangle
= 
k\left\langle\left(c-\frac{m^4}{2\lambda}+\frac{\lambda}{2}\left(x+\frac{m^2}{\lambda}\right)^2\right)^{k-1}\right\rangle -B_k
\end{equation}
where $B_k$ is defined by eq.~(\ref{SDEq}). These identities would be quite
difficult to guess, if the only inputs were the moemnts, $\left\langle
x^p\right\rangle$, given by eqs.~(\ref{sum_roots}) or~(\ref{new_vev_x_moms}).  

The stochastic approach, therefore, helps us 
(a) to  obtain concrete expressions for
the moments,$\left\langle x^p\right\rangle$, of the ``physical'' variable,
$x$, without prior knowledge of  the probability density, $\rho(x)$, 
and (b) to deduce identities between these moments, that would be very
difficult to deduce from scratch.

An interesting question is, whether the moments are compatible with a Gaussian
distribution for the scalar. One way to address it is to compute the connected
four--point function, 
\begin{equation}
\label{4pointfconn}
\left\langle x^4\right\rangle_c\equiv 
\left\langle x^4\right\rangle-3\left\langle x^2\right\rangle^2+3\left\langle
x^2\right\rangle\left\langle x\right\rangle^2-\left\langle
x^3\right\rangle\left\langle x\right\rangle
\end{equation}
which vanishes for a Gaussian distribution as a consequence of Wick's
theorem. Using eqs.~(\ref{sum_roots}),or (\ref{new_vev_x}), 
we find, indeed, that it doesn't vanish. By dimensional analysis and scaling 
it is of the form 
\begin{equation}
\label{4pointfx}
\left\langle x^4\right\rangle_c = 
\left(\frac{m^2}{\lambda}\right)^4 f\left(\frac{\lambda}{m^4}\right)
\end{equation}
and we expect that $f(g)\to 0$ as $g\to 0$, since, for $\lambda/m^4\to 0$ the
superpotential is quadratic. Using either of the
expressions~(\ref{sum_roots}),(\ref{new_vev_x}) we find that 
\begin{equation}
\label{4pointfx1}
f\left(\frac{\lambda}{m^4}\right) = \frac{4(6-\pi)}{\pi}\left(\frac{2\lambda}{m^4}\right)^2
\end{equation}
for the case when $c=m^4/(2\lambda)$. We can, of course, 
compute it for any value of the coefficient of the linear term.
We conclude that, for any, finite, value of $\lambda/m^4$, the distribution of
the scalar differs markedly from that of a Gaussian, and the ``non-Gaussian''
effects become more marked with increasing $\lambda/m^4$ at 
fixed $m^2/\lambda$.  

Now let us turn to the quartic superpotential, when $m^2<0,\lambda>0$. 
As we saw, despite the presence of fermionic zeromodes, supersymmetry is
realized. However since we can't invert, efficiently, the Nicolai map, but
compute the moments of the scalar by a way that doesn't, immediately, seem
related to the auxiliary field, to show that the moments, computed this way,
do satisfy the stochastic identities is not completely obvious.  

The master identity, that expresses the fact that the auxiliary field, $F=m^2
x + (\lambda/6)x^3$, in the limit when the cutoff is removed, 
is drawn from a Gaussian distribution, reads
\begin{equation}
\label{quartic_aux}
\begin{array}{l}
\displaystyle
k\left\langle F^{k-1}\right\rangle_F = \left\langle
F^{k+1}\right\rangle_F\Rightarrow \\
\displaystyle
k\left\langle\left(-|m^2| x + \frac{\lambda}{6}x^3\right)^{k-1}\right\rangle = 
\left\langle\left(-|m^2| x + \frac{\lambda}{6}x^3\right)^{k+1}\right\rangle 
\end{array}
\end{equation}
along with the ``initial condition'', $\left\langle F\right\rangle =0$. The
master identity, for $k=1$, implies the first, non-trivial, relation 
\begin{equation}
\label{vevF2quartic}
\left\langle F^2\right\rangle_F = \left\langle 1\right\rangle_F = 1\Rightarrow
\left\langle\left(-|m^2| x + \frac{\lambda}{6}x^3\right)^2\right\rangle = 1
\end{equation}
We shall start with the 1--point function for the auxiliary field:
Supersymmetry implies that $\left\langle F\right\rangle_F = 0$, therefore,
\begin{equation}
\label{1pointFquartic}
-|m^2|\left\langle x\right\rangle + \frac{\lambda}{6}\left\langle x^3\right\rangle = 0
\end{equation} 
Once more it's useful to work with rescaled variables. Let us set 
\begin{equation}
\label{rescaled_quartic}
x\equiv X\left(\frac{2|m^2|}{\lambda}\right)^{1/2}
\end{equation}
The moments then are calculated through 
\begin{equation}
\label{quartic_moms_rescaled}
\left\langle x^p\right\rangle = \left(\frac{2|m^2|}{\lambda}\right)^{p/2}
\frac{1}{Z}\int_{-\infty}^\infty
dX\,(-1+X^2)X^p\,e^{-\frac{|m^2|^3}{\lambda}\left(-X+\frac{X^3}{3}\right)^2}
\equiv \left(\frac{2|m^2|}{\lambda}\right)^{p/2}
\frac{\left\langle (-1+X^2)X^p\right\rangle}{\left\langle (-1+X^2)\right\rangle}
\end{equation}
The scaling relation for the moments of the quartic superpotential becomes
\begin{equation}
\label{scaling_quartic}
\left\langle x^p\right\rangle = \left(\frac{2|m^2|}{\lambda}\right)^{p/2} h_p\left(\frac{\lambda}{|m^2|^3}\right)
\end{equation}
and we can write the master identity~(\ref{quartic_aux}) in terms of the 
rescaled variable $X$, in the following way
\begin{equation}
\label{quartic_aux_rescaled}
\left\langle\left(-X+\frac{X^3}{3}\right)^{k+1}\right\rangle =
\left(\frac{\lambda}{2|m^2|^3}\right)^2
\left\langle\left(-X+\frac{X^3}{3}\right)^{k-1}\right\rangle =
k\left\langle\left(-X+\frac{X^3}{3}\right)^2\right\rangle
\left\langle\left(-X+\frac{X^3}{3}\right)^{k-1}\right\rangle \\
\end{equation}
with 
\begin{equation}
\label{rescaled_moms_quartic}
\left\langle\left(-X+\frac{X^3}{3}\right)^k\right\rangle = 
\frac{
\int_{-\infty}^{\infty}dX\,(-1+X^2)\left(-X+\frac{X^3}{3}\right)^k\,
e^{-\frac{|m^2|^3}{\lambda}\left(-X+\frac{X^3}{3}\right)^2}
}
{
\int_{-\infty}^{\infty}dX\,(-1+X^2)\,
e^{-\frac{|m^2|^3}{\lambda}\left(-X+\frac{X^3}{3}\right)^2}
}
\end{equation}
The relation, $\left\langle F\right\rangle_F=0$ becomes
\begin{equation}
\label{1pointFquartic_resc}
-\left\langle X\right\rangle + \frac{\left\langle X^3\right\rangle}{3}=
\left\langle -X + \frac{X^3}{3}\right\rangle = 0
\end{equation}
If we write it out:
\begin{equation}
\label{pointFquartic_resc1}
\int_{-\infty}^\infty dX\,\left(-X +
\frac{X^3}{3}\right)\,\left(-1+X^2\right)
\,e^{-\frac{|m^2|^3}{\lambda}\left(-X+\frac{X^3}{3}\right)^2}
= 0
\end{equation} 
it becomes obvious, since, in this guise, the integral exists, thanks to the
exponential factor and the integrand is the product of an even function and an
odd function, integrated along an interval, symmetric about the origin, 
therefore it vanishes. Indeed, any odd moment vanishes as a result of the
symmetry, $X\leftrightarrow -X$.

Let us look at a, possibly,  less trivial relation, namely $\left\langle
F^2\right\rangle_F = 1$. In terms of the $\left\langle X^p\right\rangle$, it
reads
\begin{equation}
\label{2pointFquartic_rescal}
\left\langle\left(-X+\frac{X^3}{3}\right)^2\right\rangle =
\frac{\lambda}{2|m^2|^3}\Leftrightarrow
\frac{\int_{-\infty}^\infty
dX\,\left(-1+X^2\right)\left(-X+\frac{X^3}{3}\right)^2\,e^{-\frac{|m^2|^3}{\lambda}\left(-X+\frac{X^3}{3}\right)^2}
}
{
\int_{-\infty}^\infty\,dX\left(-1+X^2\right)\,^{-\frac{|m^2|^3}{\lambda}\left(-X+\frac{X^3}{3}\right)^2}
}
= \frac{\lambda}{2|m^2|^3}
\end{equation}
If we could perform the change of variables,
$F\equiv\sqrt{|m^2|^3/\lambda}\left(-X+(X^3/3)\right)$, it would be
obvious. The analysis in the previous section confirms this expectation. Let
us show that it is possible to recover this result by computing the moments
according to eq.~(\ref{quartic_moms_rescaled}). 

\begin{figure}[thp]
\begin{center}
\includegraphics[scale=1.2]{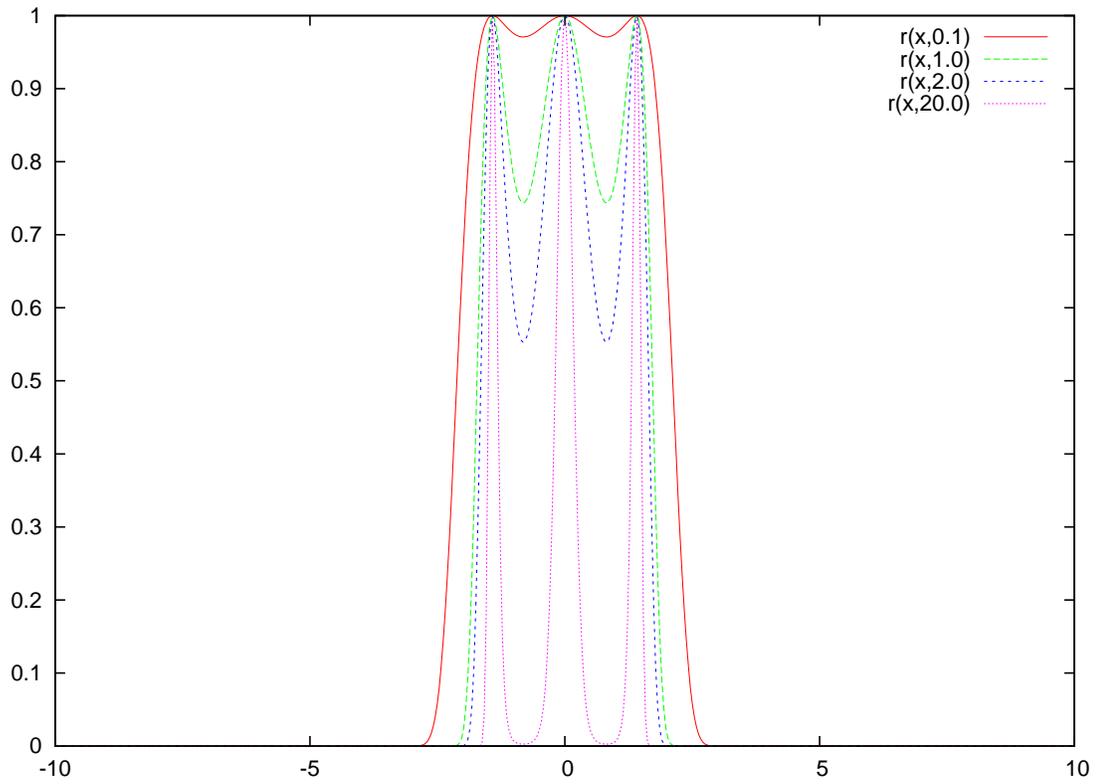}
\end{center}
\caption[]{$\exp\left(-(|m^2|^3/\lambda)\left(-X+(X^3/3)\right)^2\right)$ as a
  function of $X$ for different values of $|m^2|^3/\lambda$.}
\label{rho_quartic}
\end{figure}

If we draw $\exp\left(-(|m^2|^3/\lambda)\left(-X+(X^3/3)\right)^2\right)$ for
different values of the coupling constant, $\lambda/|m^2|^3$,
cf.~fig.~\ref{rho_quartic}
that cover the interval from ``weak'' ($\lambda/|m^2|^3<1$) to ``strong''
($\lambda/|m^2|^3 > 1$) coupling, we note that, for weak coupling, the density is
concentrated around the ``classical'' minima, while, as the coupling
increases, it becomes easier to explore other values, in particular around the
local maxima of the ``classical action''--however we do note that,
for $|X|>5$, the density becomes negligible: for $|X|\approx 5$ the density is
smaller than $\sim\exp(-37^2|m^2|^3/\lambda)$. Therefore we may cutoff the
integrals at $|X|=5$. 

\begin{figure}[thp]
\includegraphics[scale=1.0]{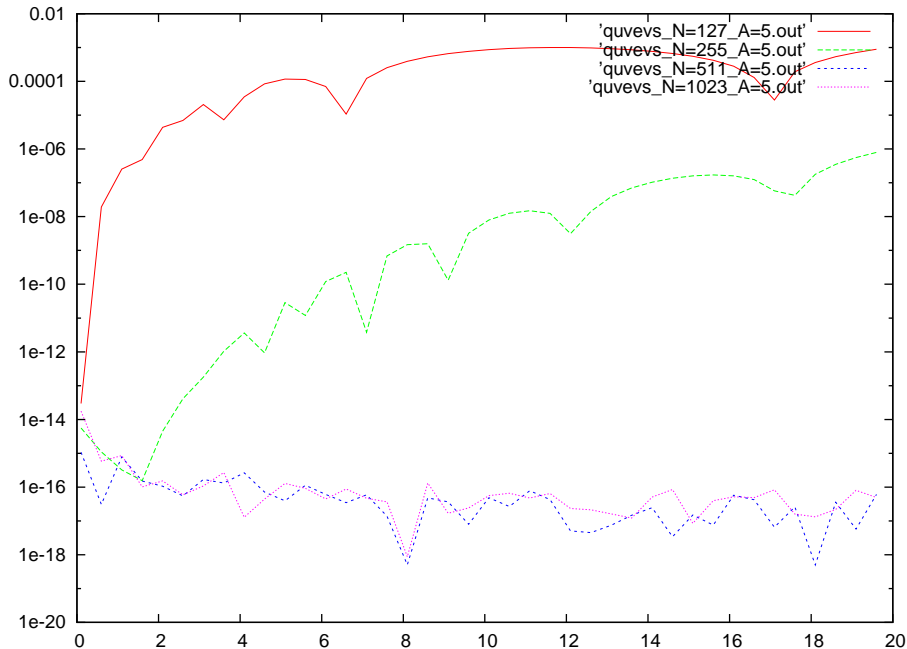}
\caption[]{Absolute value of the difference of the two sides
  of relation~(\ref{2pointFquartic_rescal}) as a function of
  $|m^2|^3/\lambda$. The vertical scale is logarithmic. 
The different curves correspond to different values of
  points for Simpson's rule, used for evaluating the integrals, in the
  interval $-5\leq X\leq 5$. }
\label{rho_quartic_vevF2}
\end{figure}

In fig.~\ref{rho_quartic_vevF2} 
we display, for different values of
$\lambda/|m^2|^3$, the absolute value of the difference between the left hand side and the right hand side of
eq.~(\ref{2pointFquartic_rescal}). We find, once more, a result that is consistent
with zero, to machine precision, when numerical integration errors become negligible. Our code passes the test. 

\newpage
\section{Conclusions and Perspectives}\label{conclusions}
We have studied the partition function of 
probabilitiy distributions of one variable,  
that are generated by solutions of Langevin's equation.  This defines a
zero--dimensional field theory. We have shown that the induced probability
distribution enjoys worldpoint supersymmetry, that can be broken by surface
terms, induced by fermionic zeromodes. The ``fermions'' enter through the
noise term. 

While a linear term in the
superpotential, also, contributes surface terms, in fact it cannot cancel
their contribution. In this way we provide concrete
examples for the boundary terms studied by Parisi and
Sourlas~\cite{parisi_sourlas}. 

Nevertheless, this breaking doesn't imply that supersymmetry isn't useful: as
we have shown, even broken, it, still imposes relations on the moments 
and  all higher moments of the auxiliary field 
are determined by its first and second moments. So the auxiliary field is
still drawn from a Gaussian distribution, ``deformed'' in an interesting way,
due to the interactions.

For practical purposes, we can find examples, where 
supersymmetry is spontaneously broken, but the goldstino is ``hard to find'',
since the average value of the auxiliary field is not the ``typical'' value,
that is drawn from its distribution. 
 
The scalar field, on the other hand, isn't drawn from a Gaussian
distribution--but we can compute its moments, at any value of the coupling, 
 quite explicitly and they determine its distribution (its effective action, in the field theory context), in principle. We have determined exact expressions
for the moments of the cubic superpotential by two independent methods:
inverting the Nicolai map and by establishing the relation to the moments of
the scalar part of the classical action. It is this second approach that most
naturally generalizes to higher dimensions, where numerical simulations are
the natural tool, which we tested on the quartic superpotential, where
supersymmetry is realized, whatever the sign of the quadratic term of the
superpotential. 

So the situation is by no means ``all or nothing'': supersymmetry either 
is realized and constrains everything or is broken and is not useful, 
but is much more interesting. The information that has  not been appreciated  
lies in the identities between moments or correlation functions. 
While in the ``traditional'' approach to supersymmetric 
field theory, one starts from the ``physical'' (albeit classical) fields and 
the auxiliary fields are, indeed, quite hard to find~\cite{dAFFV}, 
in the stochastic approach the auxiliary fields are the starting point and the
 ``physical'' fields are ``emergent'' quantities. 
In this approach supersymmetry is a natural symmetry. 
So to understand it better, it might seem that the stochastic 
approach, through the access it provides to the moments/correlation functions,
might prove much more useful. 

In fact a natural direction of further study is the following: The reason why
the auxiliary field is drawn from a Gaussian distribution is that it was,
initially, at least, identified with the noise in the Langevin equation. 
So we might consider other stochastic equations, where the noise is drawn 
from other distributions, e.g. ``colored'' noise~\cite{stochSUSY} and, of
course we can imagine further generalizations. These might lead to more
general supersymmetry-breaking patterns, that definitely deserve to be studied
in higher dimensions. 
  
The most straightforward generalization is to ``worldline supersymmetry'',
i.e. supersymmetric quantum mechanics~\cite{witten,windey,waseda}. This has
been, mainly, studied in a way that stresses its similarity to 
(non-relativistic) quantum mechanics rather than quantum field theory (an
exception is ref.~\cite{windey}, where, however, the focus was on topology and
not on the stochastic identities themselves, that were studied in
ref.~\cite{dAFFV} in an approach that combined canonical and path integral
approaches ). 
Lattice studies~\cite{lattice_WZ} use the Nicolai map but,
apparently, do not consider the stochastic identities themselves. 
  These are ``broken'' by
the lattice since the auxiliary field has a local propagator, that becomes
ultra-local only in the continuum limit (at least for the free theory). 

In this paper we have tried, on the contrary, to set up the formalism that 
most readily generalizes
to quantum field theory, namely the (zero--dimensional analog of the) path
integral formalism and the WT identities between moments.  
We have thereby elucidated the role of the distribution of the auxiliary field
in characterizing supersymmetry breaking and developed the tools that will be
useful for the higher dimensional cases. 

Our results can be considered as describing the degrees of freedom that live
on the boundary of Euclidean time, with  supersymmetric quantum mechanics 
describing the ``bulk'' theory. For the cubic superpotential, supersymmetry is 
therefore broken both in the 
bulk~\cite{parisi_sourlas,cecotti_gir,damgaard,witten,vanholten_etal} and 
on the boundaries. For the quartic superpotential it is broken only in the
bulk. These statements, however, do not take into account the coupling between 
bulk and boundaries, which is the issue that remains to be studied (work in this
direction, in the canonical formalism, was done in ref.~\cite{waseda}).

We have studied the case of one chiral multiplet and a natural generalization
is to $N_\mathrm{f}>1$ chiral multiplets--which also lead naturally to
theories with extended supersymmetry. 
In that case the technical
difficulty stems from the fact that we must ``excise'' not just a point but a
vector space, generated by the zeromode(s) of the Jacobian,
$N_\mathrm{f}\times N_\mathrm{f}$,  matrix. The use of collective
coordinates~\cite{gervais_sakita} is the natural tool, already at this level.

Finally, another topic concerns supersymmetric gauge theories. Here the
problem is how to handle gauge invarince. In  stochastic
quantization~\cite{parisi_wu} gauge fixing is not mandatory, but only gauge
invariant quantities are well-defined when the equilibrium limit is taken. If
we work on the lattice, with gauge links that take values in a compact group,
all observables are gauge invariant, but the gauge links are constrained:
e.g. they are unitary matrices, for unitary groups.
 We may solve these constrants, however~\cite{nicolis10},
so it will be interesting to study the corresponding Langevin equation and 
whether we may obtain supersymmetric gauge theories this way. 

For, indeed, the question remains, how ``target space'' supersymmetry can
be obtained by the stochastic approach. Parisi and Sourlas did succeed in
obtaining the ${\mathcal N}=2$ Wess-Zumino model in two spacetime dimensions,
but found an obstruction in four dimensions. de Alfaro, Fubini , Furlan and
Veneziano~\cite{dAFFV} obtained some, partial, results, in particular for
gauge theories, but the issue remains
open, how to write a lattice theory in this formulation. This is one reason
why the ``orbifold'' approach~\cite{orbifold} has since been used. It is,
therefore, of interest to test its assumptions by another formulation.

{\bf Acknowledgements:} We acknowledge the helpful interactions within the
Master program ``Mod\`eles de Physique Non-lin\'eaire'' at the Physics
Department at Tours that fostered this project. SN would like to thank
B. Boisseau, H. Giacomini and A. Niemi for discussions and, especially,
J. Iliopoulos for bracing exchanges.

\end{document}